# Epitaxial graphene on SiC: Modification of structural and electron transport properties by substrate pretreatment

**Mattias Kruskopf**[1], **Klaus Pierz**[1], **Stefan Wundrack**[1], **Rainer Stosch**[1], **Thorsten Dziomba**[1], **Cay-Christian Kalmbach**[1], **André Müller**[1], **Jens Baringhaus**[2], **Christoph Tegenkamp**[2], **Franz J. Ahlers**[1] and **Hans W. Schumacher**[1]

[1] Physikalisch-Technische Bundesanstalt, Bundesallee 100, 38116 Braunschweig, Germany

[2] Institut für Festkörperphysik, Leibniz Universität Hannover, Appelstraße 2, 30167 Hannover, Germany

## Abstract

The electrical transport properties of epitaxial graphene layers are correlated with the SiC surface morphology. In this study we show by atomic force microscopy and Raman measurements that the surface morphology and the structure of the epitaxial graphene layers change significantly when different pretreatment procedures are applied to nearly on-axis 6H-SiC(0001) substrates. It turns out that the often used hydrogen etching of the substrate is responsible for undesirable high macro steps evolving during graphene growth. A more advantageous type of sub-nanometer stepped graphene layers is obtained with a new method: a high-temperature conditioning of the SiC surface in argon atmosphere. The results can be explained by the observed graphene buffer layer domains after the conditioning process which suppress giant step bunching and graphene step flow growth. The superior electronic quality is demonstrated by a less extrinsic resistance anisotropy obtained in nano-probe transport experiments and by the excellent quantization of the Hall resistance in low-temperature magneto-transport measurements. The quantum Hall resistance agrees with the nominal value (half of the von Klitzing constant) within a standard deviation of $4.5 \times 10^{-9}$ which qualifies this method for the fabrication of electrical quantum standards.

# 1. Introduction

The verification of the extraordinary electronic properties of monolayer graphene sheets has launched a huge research interest in this novel material which has the potential to become a basic material for electronic devices. [1], [2], [3], [4] The appearance of the quantum Hall effect allows a first graphene-based application as a robust quantum-resistance standard for the unit Ohm with the potential to observe quantization even at room temperature. [5], [6], [7] A key point for fabrication of graphene based devices is the domain size of the two-dimensional crystal. One successful technique is the epitaxial growth of graphene by thermal decomposition of a SiC surface in argon atmosphere. The Ar pressure allows a higher process temperature which improves the surface kinetics and the formation of a long-range homogeneously $sp^2$-bonded graphene network. [8], [9] By using the Si-terminated SiC surface a single layer of graphene forms in a quasi self-limiting process on top of a so called graphene buffer layer which is still partly covalently bonded to the substrate. [10] Simultaneously, the high temperature forces an energetically driven restructuring of the SiC surface. The surface steps bunch together and form regular terrace structures depending on the process temperature [11] and substrate mis-orientation. [12] For graphene device applications, however, it is of great interest to control the SiC surface morphology. Graphene bilayer stripes forming at high terrace edges seem to be responsible for a strong anisotropy of the magneto-resistance. [13] Recently, it was found that also small isolated bilayer patches on a monolayer graphene sheet grown at 2000 °C are detrimental for its electronic properties, [14] and even the graphene doping level can be related to the surface terrace width. [15] In practice it is difficult to control all growth parameters simultaneously. On almost perfect on-axis SiC substrates and at growth temperatures as high as 2000 °C terraces of 6 μm width were found with step heights of 1 nm - 1.5 nm which allow a continuous graphene film growth across these step edges. [12] In another study several nanometer high step edges have been observed when the growth was performed at lower temperature of 1650 °C on nominally on-axis substrates. [9]

In this study we show that for substrates with very small mis-orientation and equal growth conditions the pretreatment of the SiC surface plays the decisive role for the graphene formation. The widely employed hydrogen etching of the substrate [9], [11] results in high terrace edges during graphene growth which is detrimental for the electronic properties. Whereas a pre-annealing procedure in Ar atmosphere yields smooth sub-nanometer stepped single layer graphene of nearly isotropic conductivity. High precision quantum Hall resistance measurements underline the high quality particularly with regard to quantum metrology. Pre-conditioning in argon atmosphere could thus open up a way for controlling the structure and morphology of graphene layers for device applications.

## 2. Methods

All substrates used in this study were taken from an epi-ready, semi-insulating 6H-SiC(0001) wafer with a very small nominal mis-orientation angle of -0.06° towards [1$\bar{1}$00] and 0.01° towards [11$\bar{2}$0]. Before cutting the wafer into pieces of 5 mm x 10 mm size the surface was protected with a thin layer of photoresist. The samples were individually cleaned in acetone and isopropanol before they were introduced into the reactor. The pretreatment and the subsequent graphene formation process were performed in the same inductively heated hot-wall reactor. [16] Two different pretreatment recipes of the SiC surface were applied: Annealing in argon atmosphere, or hydrogen etching in forming gas (500 sccm 5% $H_2$/Ar), both at 1 bar and 1400 °C for 30 min. For the subsequent graphene growth the sample was heated (~400 °C/min) in Ar atmosphere of 1 bar to the growth temperature of 1900 °C and kept for 5 min. The same parameters were used for all samples including a not pretreated one as reference. (Details of graphene preparation see also supplementary data.)

## 3. Results and discussion

*3.1 Morphological and structural characterization*

The atomic force microscopy (AFM) image in figure 1(a) shows the graphene layer which was grown after the SiC substrate was pre-annealed in Ar atmosphere at 1400°C. We observe a very smooth surface with a homogeneous distribution of parallel nearly straight terrace steps of about (0.8 ± 0.1) nm in height which corresponds to three SiC bilayers (half SiC unit cell). The terraces show a width of about 0.7 μm in agreement with the nominal mis-orientation angle of the wafer. Occasionally, the terraces are interrupted by small triangular shaped indentations with an angle of 120° at the top and by larger indentations elongated along the terraces. About 50% of the indentations are isolated while others occur in pairs on two successive terraces.

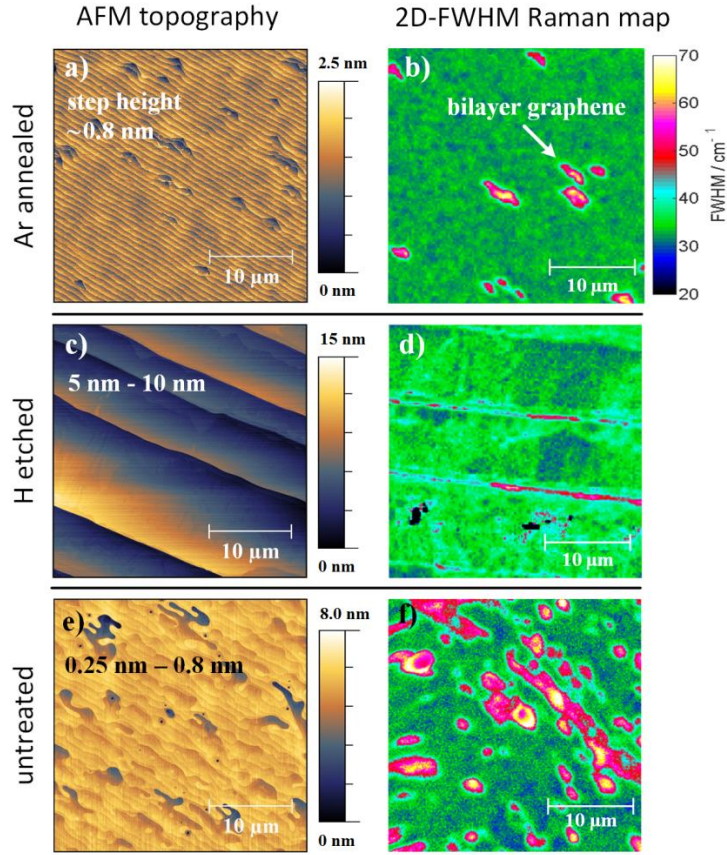

**Figure 1**. Atomic force microscopy (left column) and Raman measurements (right column) of epitaxial graphene grown at 1900 °C in Ar atmosphere on 6H-SiC (0001) substrates after different SiC surface pretreatment: (a), (b) annealing in Ar atmosphere at 1400 °C, (c), (d) H etching at 1400 °C and (e), (f) no pretreatment. (a), (c), (e) AFM topography images (30 μm x 30 μm) and (b), (d), (f) Micro-Raman maps (30 μm x 30 μm) of 2D-peak width (FWHM) indicating monolayer (blue and green) and bilayer graphene (red and yellow). AFM and Raman measurements were performed at different sites on the same sample.

Micro-Raman spectroscopic measurements of this sample show an intense 2D peak signal at wavenumbers around 2736 cm$^{-1}$ which proves the formation of graphene. From the full width at half maximum (FWHM) of the 2D peak the number of graphene layers can be estimated, irrespective of whether exfoliated or epitaxial graphene is considered. [17], [18] (Details of Micro-Raman measurements and evaluation see also supplementary data.) The 2D-FWHM topography is depicted in figure 1(b). The small FWHM values of 30 $cm^{-1}$ to 40 $cm^{-1}$ (blue and green shade) obtained from Lorentzian fitting of each 2D band indicate the presence of a fairly homogeneous and continuous monolayer graphene sheet. Only a few small domains of FWHM = 45 $cm^{-1}$ to 65 $cm^{-1}$, (red and yellow shade) in between are observed which are attributed to bilayer graphene. The size of the smaller bilayer domains agrees well with the terrace widths and they are most likely associated with the small triangular indentations observed with AFM. The larger bilayer domains are correlated with the elongated indentations as will be shown in the following. Note that micro-Raman and AFM were measured at different positions.

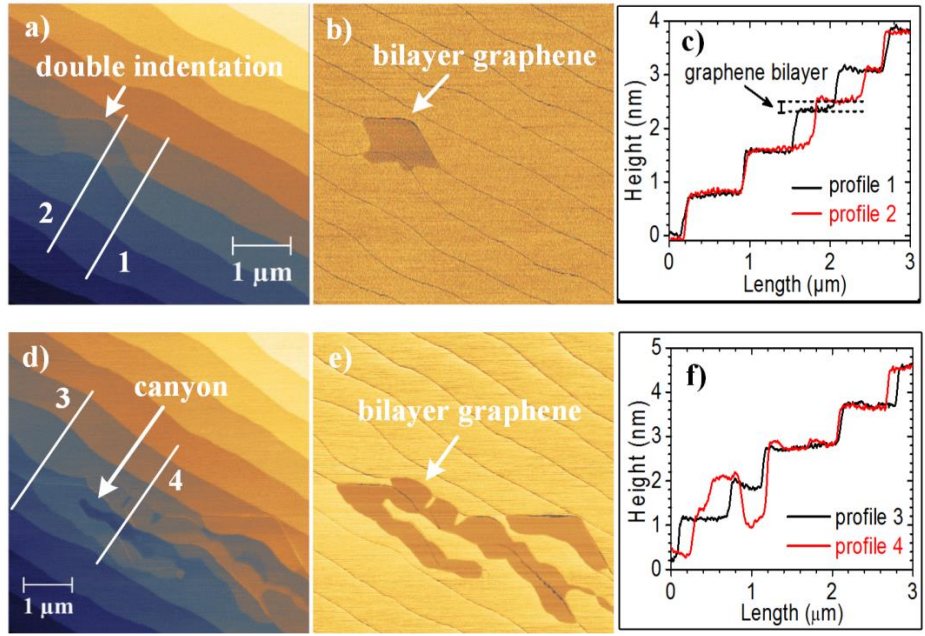

**Figure 2.** Epitaxial graphene on SiC surface pre-annealed at 1400 °C in Ar atmosphere. (a), (d) Tilted AFM images in order to obtain horizontally aligned terraces. The terrace steps have a height of ~0.8 nm. (b), (e) Corresponding phase images distinguish between monolayer graphene (yellow) and bilayer domains (brown). Uncovered SiC is not detected. (a) shows a pair of small subsequent indentations with the upper one covered with bilayer graphene. (d) shows a large elongated indentation of canyon-like type surrounded with bilayer domains. (c), (f) Height profiles 1, 2, 3, 4 along the lines in (a) and (d), respectively.

Figure 2(a) displays the tilted topographical AFM image of two small successive indentations. This presentation reveals the homogeneity of the surface with regular step heights of (0.8 ± 0.1) nm and that the indentations are concave erosions in the terrace plane. The contrast change in the material sensitive AFM phase image in figure 2(b) indicates that in this special case only the upper indentation is covered with bilayer graphene. From the comparison of the line profiles 1 and 2 plotted in figure 2(c) a height of the graphene bilayer of (0.25 ± 0.1) nm is estimated in good agreement with [19]. Other double indentations show bilayer domains in each indentation. The reason for the formation of the small indentations is not clear, but might be related to crystallographic defects or step height variations. [20] Figure 2(d) shows an example for the elongated canyon-like indentations which differ from the small ones not only in size and shape. The phase image in figure 2(e) reveals that here the bottom of the canyon is covered with monolayer graphene and bilayer graphene has formed around it. The comparison of the line profiles 3 and 4 plotted in figure 3(f) show that the canyon walls are higher than regular terrace steps. On the lower side of the canyon a height of about 1 nm is estimated and 1.5 nm for the step to the upper terrace. The formation of multilayer graphene at higher terrace edges is a general observation in the literature. [9] We can state that two kinds of bilayer graphene domains have formed at topographically different sites on this type of shallowly stepped graphene layer.

The sample which was etched in hydrogen at 1400 °C before graphene growth reveals a very different surface structure. The AFM image in figure 1(c) shows broad macro-terraces of 3 µm to 10 µm widths and macro-steps of 5 nm to 10 nm in height which is a result of giant step bunching of the SiC surface. The Raman 2D-FWHM map in figure 1(d) shows monolayer graphene coverage on the terraces and small bilayer stripes located at the edges. [9] Although the graphene was grown with the same parameters on the same kind of SiC substrates the surface morphology and the graphene structure is significantly different, demonstrating the strong impact of the pretreatment procedure.

The importance of the surface pretreatment on the graphene growth becomes obvious when the pretreatment is omitted and the sample is directly heated to 1900 °C. The AFM image in figure 1(e) shows a very inhomogeneous and rough surface covered with many canyon-like indentations. No giant step bunching is observed and the terrace width varies between 0.2 µm and 1 µm with step heights between 0.25 nm and 0.8 nm and higher steps at the canyon-like defects. The sample is reminiscent of the Ar pre-annealed one but with a much lower degree of order. The 2D-FWHM map in figure 1(f) shows that the surface is significantly covered by graphene bilayer domains and finally underscores the rising inhomogeneity without any sample pretreatment. These results show that Ar and H pretreatment improve the homogeneity of the graphene sheet in a specific way which opens up a route to control the growth of the desired type of graphene. But how does the pretreatment procedure modify the SiC surface that it leads to such different kinds of graphene? The smooth SiC surface of the reference sample before graphenization is shown in the AFM image in figure 3(a). Figure 1(e) demonstrates that a fast temperature ramp up to 1900 °C prevents a regular terracing of the SiC surface and that also the giant step bunching is suppressed. As a consequence an inhomogeneous graphene layer is formed. Note that surfaces with larger mis-orientation angles can undergo giant step bunching without H pre-etching. [11] In another study graphene layers with 1 nm - 1.5 nm steps were grown without Ar pretreatment [12] which might be due to unknown differences in the wet chemical treatment, wafer properties or process parameters.

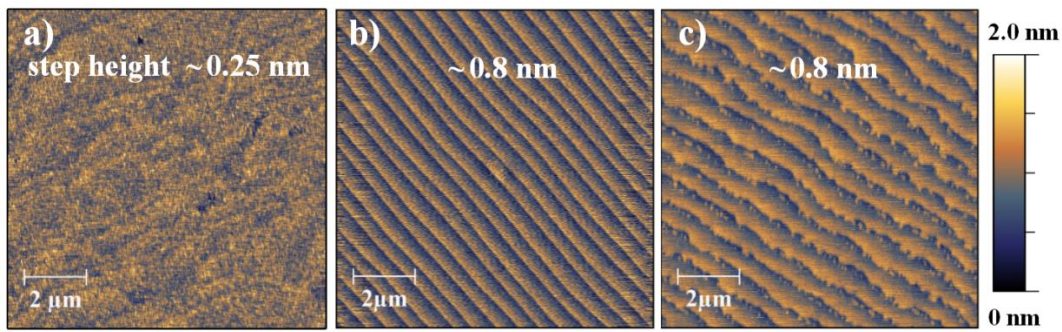

**Figure 3.** AFM images (10 µm x 10 µm) of 6H-SiC(0001) surfaces of (a) untreated reference sample (rotated 90° cw), (b) H etched at 1400 °C and (c) annealed in Ar atmosphere at 1400 °C before graphene growth.

In contrast, the hydrogen etching of the SiC surface at 1400 °C leads to a very homogeneous terracing (width ~0.5 µm) with parallel and equidistant edges (~0.8 nm in height) as can be seen in figure 3(b). This is explained by surface energy minimization among different SiC bilayers.

[21] When the temperature is further increased the pre-stepped surface undergoes the giant step bunching [9], [11] with elevated step heights which favor a step flow growth of graphene. [19] This leads to high quality graphene layers on the terraces accompanied by multilayer graphene stripes at the terrace edges [9] as can be seen in figure 1(b).

Also the Ar annealed sample in figure 3(c) shows a significant restructuring of the SiC surface compared to the pristine surface in figure 3(a). A step bunching process has lead to a nearly regularly stepped surface with narrow terraces (width of ~0.7 µm) and shallow microsteps (~0.8 nm in height), slightly wider than after hydrogen pretreatment. But unlike in the hydrogen case, the terrace edges are not straight and there are small islands on the terraces. At the first glance the surface topography is not very different compared to the H etched surface in figure 3(b) which leads to the question why at higher temperatures a significantly different graphene structure evolves.

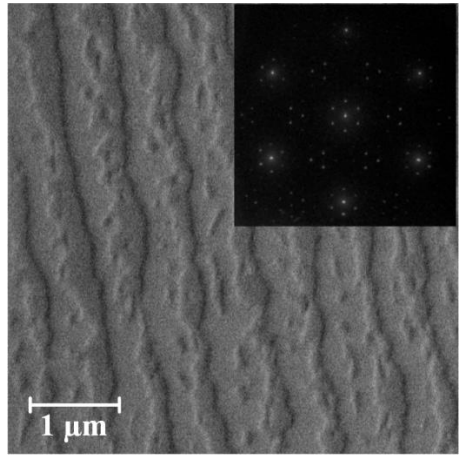

**Figure 4.** SEM image (5 µm x 5 µm) taken at 1 keV of a SiC surface after annealing in Ar atmosphere at 1400 °C. Dark areas are attributed to buffer layer domains and light grey to SiC areas. Inset shows 120 eV LEED image of the same surface showing a (6√3 × 6√3) diffraction pattern typical for graphene buffer layer coverage.

A crucial difference between the surfaces of both SiC substrates was observed in low-energy electron diffraction (LEED) measurements. The H etched surface shows a (1×1) pattern (not shown) which indicates an unreconstructed bare SiC surface. The Ar annealed surface, however, shows a (6√3 × 6√3) buffer layer pattern, [10] see inset of figure 4. This means that already at 1400 °C under Ar atmosphere Si sublimation occurred and that during 30 min the remaining carbon atoms formed extended buffer layer domains. They are visualized by scanning electron microscopy (SEM) and can be seen in figure 4 as dark ribbons along the terrace edges whereas the SiC surface appears brighter. (Details of SEM measurements see also supplementary data.) Small islands which were also observed in the corresponding AFM image are attributed to buffer layer islands on top of the SiC surface. We propose the extended buffer layer ribbons as the reason for the different surface structure of the Ar pre-annealed graphene: They stabilize the SiC surface and the terrace structure is "pinned" in the current state. The giant step bunching cannot occur even when the temperature is increased above 1400 °C and the narrow terraces are preserved when the subsequent graphenization process takes place. This model is in good agreement with the observation that buffer layer ribbons stop the thermal decomposition of triple SiC bilayer steps. [22] Furthermore, the observed canyons can be naturally explained by gaps in the buffer layer ribbons. There the decomposition of the SiC terrace locally proceeds until the next buffer layer domain stops the process. [22]

Note that triple SiC bilayer steps (half SiC unit cell) are the lowest step height observed in a thermal decomposition or etching process of 6H-SiC(0001) and the Ar pretreated SiC surface is therefore best suited for graphene growth. Our results suggest that the graphene formation on a shallowly stepped surface depends on three crucial points. The important starting condition is a SiC surface with a very low mis-orientation angle. Secondly, a pre-terracing of the SiC surface with shallow steps is needed which is performed in our case by annealing at 1400 °C in Ar atmosphere. And third, further giant step bunching must be suppressed when the temperature is raised for graphenization which was achieved by the formation of buffer layer stripes within the Ar pre-annealing process.

In case of the H etching the latter point is not fulfilled. Although a similar pre-terracing of the SiC surface is obtained as for the Ar pre-annealing the H etching process leaves behind a bare SiC surface without buffer layer stripes. When the temperature is raised for graphenization the SiC surface undergoes the giant step bunching process accompanied by a step flow growth of graphene which results in wider terraces and higher step edges decorated with graphene bilayer stripes. It is therefore assumed that this type of graphene is always obtained when the substrate was H etched independent of the mis-orientation of the 6H-SiC(0001) surface.

*3.2 Electrical and magneto-transport characterization*

The quality of the graphene layers on the Ar pretreated substrate is demonstrated by room temperature nano-scale electrical transport measurements with an *Omicron* UHV nanoprobe system [23] in linear four-probe configuration (inset in figure 5) which is sensitive solely to

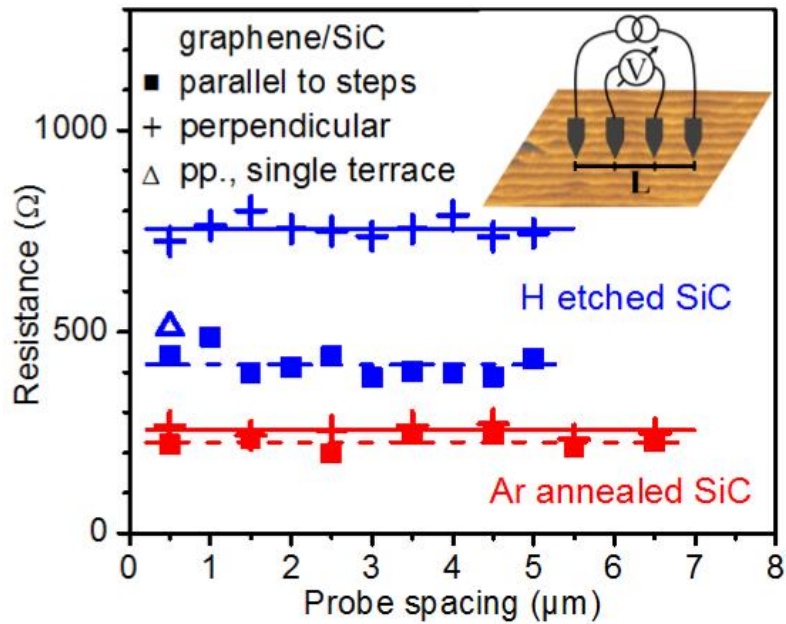

**Figure 5.** Nano-probe electrical resistance measurements ($T$ = 300 K) in linear four-probe configuration (see inset) as a function of probe spacing $L$ at graphene monolayers grown on H etched (blue) or Ar annealed (red) 6H-SiC(0001) surfaces. Measurement probes aligned parallel (∥) or perpendicular (⊥) to the terrace steps. For all perpendicular measurements at the H pre-etched sample one step edge laid between the inner probes except one measurement for $L$ = 0.5 μm with probes on the same terrace (marked as Δ). The plotted resistance values are calculated from the linear $V(I)$-curves in the ±10 μA range.

extrinsic anisotropies like step edge effects. [27], [28] The four-probe resistances, $R_∥$ and $R_⊥$, were measured parallel and perpendicular to the step edges as a function of the probe spacing $L$. The resistance values for each sample type, plotted in figure 5, are independent of $L$ which proves the 2-dimensional nature of the graphene sheets.[24] The graphene of the Ar pre-annealed sample shows values of $R_\parallel = 227\pm17\ \Omega$ and $R_\perp = 255\pm14\ \Omega$ which corresponds to a slight anisotropy of 12%. In contrast, the graphene sheet on the H pretreated substrate shows a much stronger resistance anisotropy of 80% ($R_\parallel = 418\pm32\ \Omega$ and $R_\perp = 756\pm24\ \Omega$) although the current has to cross only one terrace step compared to up to nine for the Ar pretreated sample. The strong impact of the step edge on the resistance is underlined by the perpendicular measurement on a single terrace ($L = 0.5\ \mu m$) which leads to a much lower value of $R_\perp = 511\ \Omega$ comparable to the parallel measurements. It shows that the step edge in the sample with H etched substrate is responsible for the large resistance anisotropy of about 80% which is comparable to other

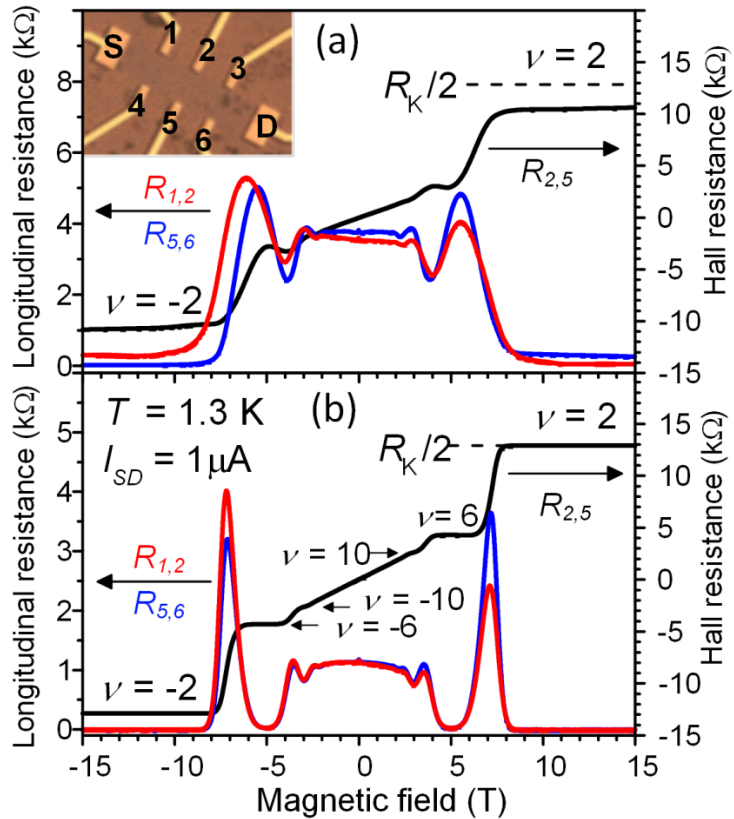

**Figure 6.** Longitudinal (red and blue curves) and Hall resistance (black curves) of monolayer graphene samples grown in Ar atmosphere at 1900 °C on 6H-SiC(0001) substrates which were pretreated by (a) H-etching and (b) by annealing in Ar atmosphere at 1400 °C. The electron densities were adjusted by a standard photo-chemical gating method: a) $n = 9.9\times10^{11}\ cm^{-2}$, $\mu = 4260\ cm^2/Vs$ and b) $n = 7\times10^{11}\ cm^{-2}$, $\mu = 7700\ cm^2/Vs$ at $T = 1.3\ K$. The inset shows a microscopy image of a typical lithographically patterned

Hall bar (100 μm x 400 μm) The dark brown spots are artifacts from the rear of the substrate. The Hall bars are aligned parallel to the step edges.

reported experimental values. [23], [24] The much smaller anisotropy for the Ar pre-annealed graphene sample shows that the origin of increased resistivity discussed in the literature, namely increased electron scattering near the terrace step edges caused by Si atom accumulation [25], defects [26], or metallic bilayer stripes [14] are not relevant for this type of graphene. It is noticeable that the absolute resistance values for the H pretreated sample are about twice as high as for the Ar pre-annealed one when we consider only parallel measurements which might be due to different electron densities or mobilities.

Finally the quality of the graphene layers were investigated by low-temperature magneto-transport measurements of a lithographically patterned Hall bar (100 μm x 400 μm) aligned parallel to the step edges, see inset in figure 6(a). In order to make the Shubnikov-de Haas (SdH) oscillations visible in the accessible range of magnetic fields the electron density was reduced by a standard photo-chemical gating method. [27] The magneto-transport measurements of the H pretreated sample (figure 6(a)) reveal a non-vanishing longitudinal resistance at a filling factor of $\nu = 2$ and a 20% lower Hall resistance value compared with the expected value of $R_K/2 = h/2e^2$ (von Klitzing constant $R_K$, Planck's constant $h$, electron charge $e$) which is attributed to parallel current paths between source and drain contacts via graphene bilayer stripes along the terrace edges.

In contrast, the magneto-transport measurements of the graphene sample on the Ar pre-annealed substrate in figure 6(b) show symmetrical longitudinal and transversal resistance curves in the measured range between -15 T and 15 T. The longitudinal resistance vanishes at filling factor $\nu = \pm 2$ for measurements at both sides of the Hall bar (red and blue curve) which indicates a good and homogeneous quantization of the relatively large graphene device. Correspondingly, the Hall plateaus at $\nu = \pm 2$ and $\pm 6$ are fully developed and even $\nu = \pm 10$ can be observed. The excellent quantization shows that the small bilayer spots observed on the graphene sheet have no detrimental effect like source-drain shorts, edge current shorts or monolayer constrictions as observed in [28].

A high precision quantized Hall resistance measurement was performed by comparison to a calibrated 100 Ω reference resistor via a cryogenic current comparator. [29] For the resistance at the ν = 2 resistance plateau ($B$ = 10 T) measured at the middle contact pair (2, 5), $T$ = 0.3 K and $I$ = 20 µA an exact quantization is obtained, agreeing with the nominal value of $R_K/2$ within a standard deviation of $4.5\times10^{-9}$. This accuracy is typical for state-of-the-art graphene (and GaAs) devices [7], [30] and it is limited in our setup by the temperature drift of the reference resistor. The result shows that Hall devices from shallowly stepped graphene layers can be used as resistance standards and underlines the superior quality of graphene grown on Ar pre-annealed 6H-SiC(0001) surfaces.

## 4. Summary and conclusions

In this work we compare epitaxial graphene grown on differently pretreated 6H-SiC(0001) substrates by AFM, Raman and electrical measurements. Although the same kind of substrate and identical growth parameters were used very different types of graphene and surface morphologies were obtained. Our results clearly show that the SiC surface plays a decisive role for the epitaxial graphene growth process and that the morphology and the type of graphene can be controlled by a suitable pretreatment of the SiC substrate. The samples with conventional hydrogen etched substrates undergo giant step bunching during graphene growth and the step edges are decorated with bilayer stripes. An annealing in argon atmosphere before graphene growth, however, results in a high quality monolayer graphene film on a surface with sub-nanometer steps. This is due to the formation of buffer layer stripes during the intermediate argon annealing step at 1400 °C which suppresses the giant step bunching. Two types of small bilayer defects are identified by AFM and Raman measurements: Small triangular shaped bilayer domains located in terrace indentations and elongated bilayer domains around canyon-like indentations. The quality of the graphene layers on argon pretreated SiC substrates is demonstrated by small extrinsic resistance anisotropy and by high precision QH resistance quantization proving the suitability of these graphene samples for electrical quantum metrology applications.

**Acknowledgements**

We thank Peter Hinze for SEM measurements. This research has been performed within the EMRP-project SIB51, GraphOhm. The EMRP is jointly funded by the EMRP participating countries within EURAMET and the European Union. One author (M. K.) is member of Braunschweig International Graduate School of Metrology (IGSM).

**Supplementary data**

# Epitaxial graphene on SiC: Modification of structural and electron transport properties by substrate pretreatment

**Mattias Kruskopf[1], Klaus Pierz[1], Stefan Wundrack[1], Rainer Stosch[1], Thorsten Dziomba[1], Cay-Christian Kalmbach[1], André Müller[1], Jens Baringhaus[2], Christoph Tegenkamp[2], Franz J. Ahlers[1] and Hans W. Schumacher[1]**

[1] Physikalisch-Technische Bundesanstalt, Bundesallee 100, 38116 Braunschweig, Germany

[2] Institut für Festkörperphysik, Leibniz Universität Hannover, Appelstraße 2, 30167 Hannover, Germany

**Graphene Growth**

Samples from the same wafer are compared in this study. Moreover, other samples also from other SiC wafers have been tested. A total number of 23 samples showed a shallow stepped surface after graphene growth on Ar pretreated SiC substrates. All these samples are taken from wafers with a very low nominal mis-orientation of (-0.06°/0.01°, -0.01°/-0.01°, 0.01°/0.06° towards major [1$\bar{1}$00]] / minor flat [11$\bar{2}$0]). Although, the two samples from the latter wafer (miscut of 0.01°/0.06°) showed no homogenous morphology. The results are similar for graphene growth at 1900 °C, 1850 °C, 1800 °C.

$H_2$ pretreatment of the SiC surface results in giant step bunching after graphenization which was confirmed by corresponding growth runs with samples from the four mentioned wafers.

The formation of shallowly stepped surfaces by Ar pre-annealing seems to be a very sensitive process which depends on the chemical cleaning of the SiC substrates as well as on the state of the reactor. We used chemical-mechanical polished (CMP) wafer with an epi-ready coating from II/VI Deutschland GmbH. Prior to cutting the SiC wafer (Automatic dicing saw DISCO DAD 3220 and diamond resin bond blades) the surface was protected by a thin layer of photoresist. An ultra-sonic cleaning in propanol and aceton of the individual samples seems to be sufficient in order to obtain the described type of shallowly stepped graphene layers. It has turned out that a further wet-chemical cleaning of the substrates by RCA clean (standard clean 2 or standard clean 1+2) or by oxygen-plasma treatment is counterproductive. This leads to giant step bunching during the graphenization process.

Moreover, the condition of the reactor plays also an important role in order to obtain graphene on shallow stepped terraces. No shallowly stepped surfaces were obtained when $H_2$ gas was used in the reactor prior to Ar pretreatment and graphene growth. At least one dummy graphenization run is necessary to get rid of residual effects. Other particle related deteriorations of the shallow stepped surface were observed.

**Micro-Raman measurements**

Raman spectroscopic measurements were acquired to examine the structural quality of the graphene sample using a *LabRAM Aramis* Raman spectrometer (*Horiba*) operated in backscattering mode. The spatial Raman imaging across an area of (30 x 30) µm² was performed with a frequency-doubled Nd:YAG-Laser with a wavelength of 532 nm, a 100x objective (NA 0.95) and a piezo sample stage with a step size of 0.2 µm, whereat the laser power was kept very low to avoid local heating of the graphene sample. A total number of 14,520 Raman spectra were recorded from this area. The laser spot diameter is about 1 µm. The effective lateral resolution is estimated to about 0.8 µm due to oversampling and the intensity profile of the spot.

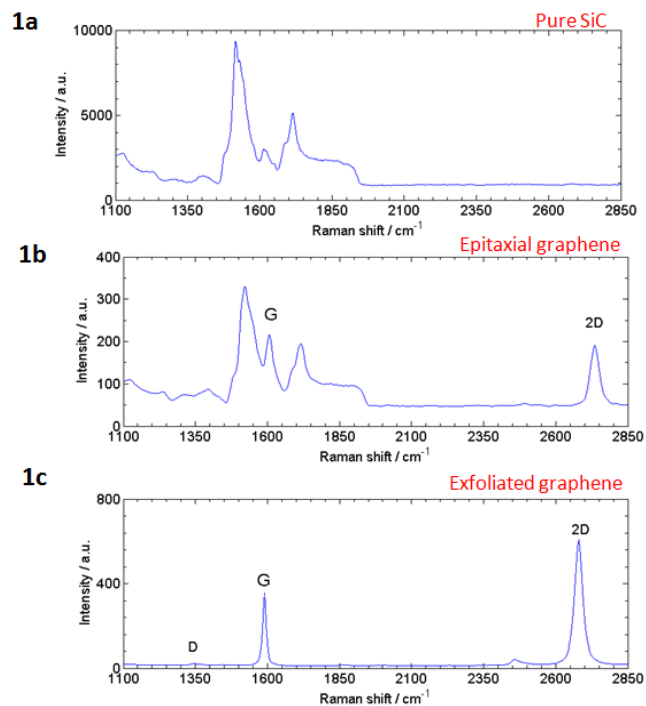

**Figure 1.** Raman spectra of (a) pure SiC substrate, (b) an epitaxial graphene layer and (c) an exfoliated graphene sheet.

Figure 1(b) shows a single Raman measurement of a shallowly stepped graphene sample. The graphene related G line close to 1600 cm$^{-1}$ and the 2D line around 2750 cm$^{-1}$ is clearly visible.[1] A broad SiC overtone as can be seen in the spectrum of pure SiC (figure 1(a)) overlaps the G peak in the range between 1500 cm$^{-1}$ and 1900 cm$^{-1}$ which unfortunately complicates the evaluation of its peak properties by Lorentzian curve fitting.[1] Additionally, no defect peak (D peak) around 1350 cm$^{-1}$ appears (1(b), 1(c)).

Since the 2D peak is not affected by the broad SiC overtone signal these peaks are evaluated by Lorentzian curve fitting. The full width at half maximum (FWHM) of the 2D peak is extracted and plotted in the 2D-FWHM topography maps because this value is related to the number of graphene layers.[1] Figure 2 gives an overview of the Raman 2D-FWHM topography map (30 μm x 30 μm) of an epitaxial graphene layer grown at 1900 °C in Ar atmosphere on four different positions (A-D), which are denoted by red dots. In positions A and B the FWHM value of 31 cm$^{-1}$ is attributed to single layer graphene , whereas the positions C and D with FWHM of 65 cm$^{-1}$ and 61 cm$^{-1}$ are attributed to bilayer graphene according to [1].

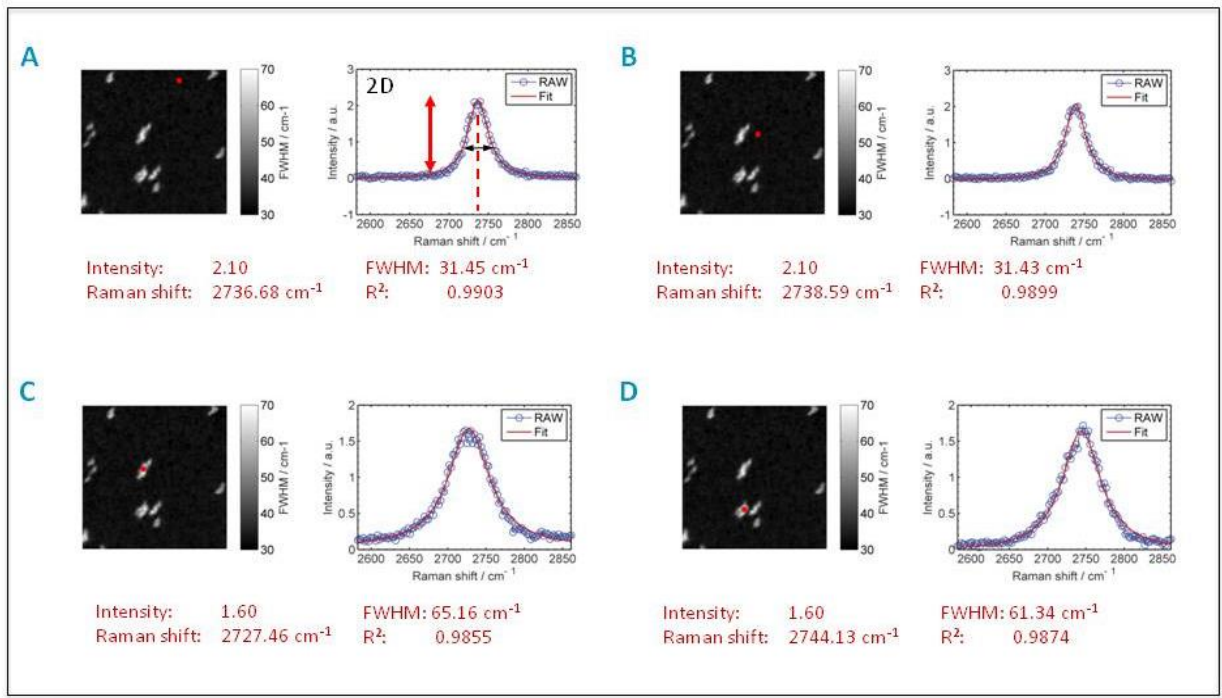

**Figure 2.** Typical examples of Lorentzian curve fitting of the 2D peaks for an epitaxial graphene layer grown on an Ar pre-annealed substrate. The red dots in 2D-FWHM maps denote the measurement position.

The total FWHM distribution of epitaxial graphene grown at 1900 °C in Ar atmosphere as a function of the 2D peak position is shown in figure 3. The scatter plot and the FWHM bar chart clearly shows that the majority of 2D linewidths are scattered around 34 cm$^{-1}$ with a standard deviation of 4 cm$^{-1}$, which emphasizes that this sample is mainly covered with monolayer graphene. A very small number of 2D peaks shows FWHM values larger than 40 cm$^{-1}$. An exact assignment of the values to bilayer graphene is difficult because of the lateral resolution of the measurement. For a conservative estimate of the area covered with bilayer graphene we attribute FWHM values larger as 45 cm$^{-1}$ to bilayer graphene.

4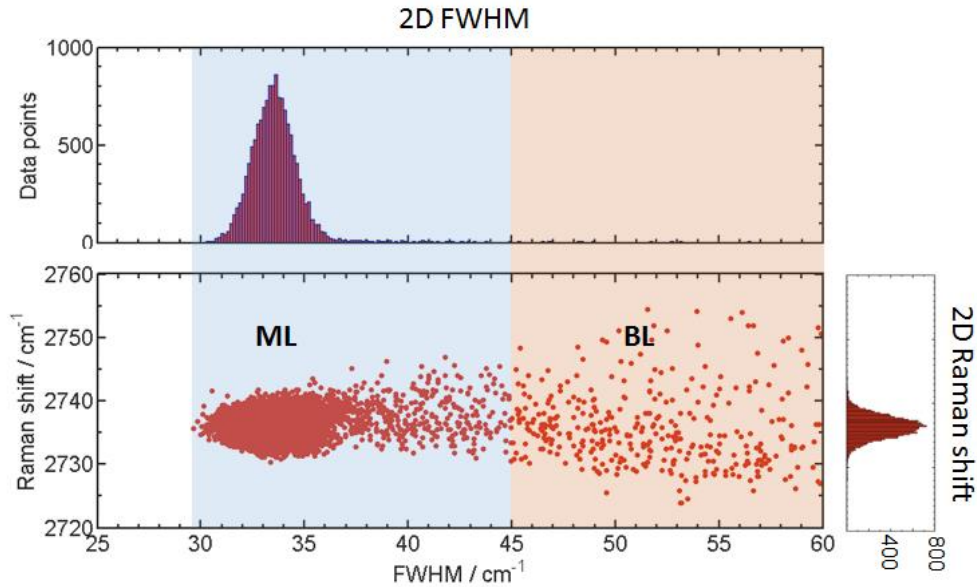

**Figure 3.** Histogram of the FWHM values and its distribution as a function of the 2D peak position for the 14520 data points from the 30 μm x 30 μm topographic map of the shallowly stepped graphene layer.

By counting the scattered data points between 30 to 45 cm$^{-1}$ and 45 to 60 cm$^{-1}$ we estimate the approximate amount of monolayer and bilayer graphene, respectively, covering the measured sample area of (30 x 30) μm². Approximately 97 % of the sample area is covered by monolayer graphene and approximately 3 % of the area is covered with bilayer graphene. Figure 3 also reveals a wide spread of the 2D peak position from 2725 cm$^{-1}$ to 2755 cm$^{-1}$ with a mean 2D peak position of 2736 cm$^{-1}$ and a standard deviation of 2 cm$^{-1}$, which is associated with an average G peak position of 1604 cm$^{-1}$ and a standard deviation of 2 cm$^{-1}$. Both peaks are shifted to higher wavenumbers and significantly differ from the peak positions of exfoliated graphene (figure 1(b) and 1(c)). As mentioned before by [2], this effect is related to strain induced into the graphene lattice by interacting with the SiC substrate. Due to the different thermal expansion coefficients of SiC and graphene, a significant blue shift of both G and 2D peak arises in the Raman spectrum. On the other hand the blue shift of the G peak is also strongly connected to the doping level in the graphene layer [3], although the 2D peak position is rather weakly affected. The G linewidth is another important feature to assess whether strain or doping is present in the graphene layer. Unfortunately, because of SiC overtone that overlaps the G peak, the evaluation of the G peak linewidth is complicated. Subtracting the Raman spectrum of epitaxial graphene by a reference spectrum of pure SiC often produces additional noise and artifacts, which affects the line shape of the G peak. Therefore, both effects strain and doping cannot be excluded.

**SEM Measurements**

For scanning electron microscopy (SEM) we used an in-lens detector at 3.2 mm working distance and an acceleration voltage of 1 keV. With these settings we obtained high resolution images of our graphene samples. SEM images of SiC substrates after annealing in argon atmosphere at 1400°C show also areas of different contrast as can be seen in figure 4(a) in the paper. The absence of graphene related Raman signals prove that no graphene has formed. The areas of different contrast are therefore attributed to buffer layer domains (detected by characteristic LEED pattern) and uncovered SiC surface areas. The dark areas are attributed to buffer layer domains because other samples show that longer annealing times (compared to 30 min for the sample in figure 4) result in an enlargement of the dark areas. In general, the assignment is difficult. The contrast can change within minutes of SEM inspection and depends on the scan speed which is probably due to sample charging effects. Note that for the main statement in the paper (namely that buffer layer stripes have formed after annealing in Ar atmosphere) the assignment is irrelevant because the dark and the bright stripes in figure 4(a) are very similar.

**AFM measurements**

In AFM measurements it is difficult to observe the buffer layer stripes because of their very shallow height. A height of about 0.3 ±0.1 nm is observed for the stripes located at the upper edge of the terrace steps. The material sensitive AFM phase images clearly show stripes of different contrast. In figure 3(c) the buffer layer stripes are not well resolved. The small islands can be resolved because they are taller, about 0.5 ±0.1 nm. They are attributed to buffer layer domains on remaining SiC islands.